# Systematic X-ray absorption study of hole doping in BSCCO – phases

R. Müller [a,*] M. Schneider [a] R. Mitdank [a] C. Janowitz [a] R.-St. Unger [a]

A. Krapf [a] H. Dwelk [a] W. Frentrup [b] R. Manzke [a]

[a] *Institut für Physik, Humboldt-Universität, Berlin, 10115 Berlin, Germany*

[b] *BESSY G.m.b.H., 12289 Berlin, Germany*

**Abstract**

X-ray absorption spectroscopy (XAS) on the O 1s threshold was applied to Bi-based, single crystalline high temperature superconductors ($HT_c$'s), whose hole densities in the $CuO_2$ planes was varied by different methods. XAS gives the intensity of the so-called pre-peak of the O 1s line due to the unoccupied part of the Zhang-Rice (ZR) singlet state. The effects of variation of the number n of $CuO_2$ - planes per unit cell (n = 1,2,3) and the effect of La-substitution for Sr for the n = 1 and n = 2 phase were studied systematically. Furthermore the symmetry of the states could be probed by the polarization of the impinging radiation.

---

[*] Corresponding Author: Institut für Physik, Humboldt-Universität zu Berlin, Invalidenstraße 110, D-10115 Berlin, Germany. Phone: +30-2093-7767 Fax: +30-2093-7729, Email: ralph.mueller@physik.hu-berlin.de

In the cuprates the electronic states near the Fermi energy $E_F$, which are involved in the low lying excitations leading to superconductivity, are mainly due to holes in the $CuO_2$ planes. Over the last years, soft x-ray-absorption spectroscopy (XAS) has been utilized to obtain information about unoccupied states at both the O and Cu sites. In particular, polarization-dependent x-ray absorption measurements provided detailed information of their symmetry and orbital character [1]. Most of the previous doping and angular dependent XAS-studies were done on the p-type compounds $La_{2-x}Sr_xCuO_4$ [2], $YBa_2Cu_3O_{7-\delta}$ [3] and $Bi_2Sr_2CaCu_2O_8$ [4]. Here we report the current status of our investigations on the electronic structure of the bismuth-based cuprates by x-ray absorption spectroscopy. The aim of the investigations is twofold: First, the Bi-system with varying number of $CuO_2$ planes offers the opportunity to study 3D coupling effects. Even though the electronic structure in all these compounds is highly two-dimensional, there is an enormous difference in the transition temperature depending on the number of $CuO_2$ layers. Second, we are interested in the influences of doping on the number of holes and the electronic structure of the Bi-based cuprates. Unlike to other studies where the hole density was varied by the oxygen stoichiometry or by replacing the Ca atoms in between the $CuO_2$ planes, we obtained the doping by replacing trivalent Sr by divalent La, i.e. leaving Ca in between the $CuO_2$ planes unchanged.

The single crystals were grown out of the stoichiometric melt. The details of the growth of the crystals and characterization will be described elsewhere [5]. We prepared several series: Optimal doped crystals of the BSCCO-family with one to three $CuO_2$ planes per unit cell with $T_C$ of 29 K, 90 K, 110 K, resp., (see [6,7]) and doped single crystals of n = 1 and n = 2 where part of the $Sr^{2+}$ is replaced by $La^{3+}$. The x-ray absorption measurements have been carried out in the fluorescence mode at the VLS-PGM beamline at BESSY. The overall energy resolution at the O 1s absorption edge was 0.3 eV; the sample temperature was 300 K. The fluorescence light was detected at an angle of 0° (**E**||a) and 85° (**E**||c) with respect to the photon beam using a Ge detector. The vacuum in the chamber was better than $10^{-9}$ mbar during the measurements.

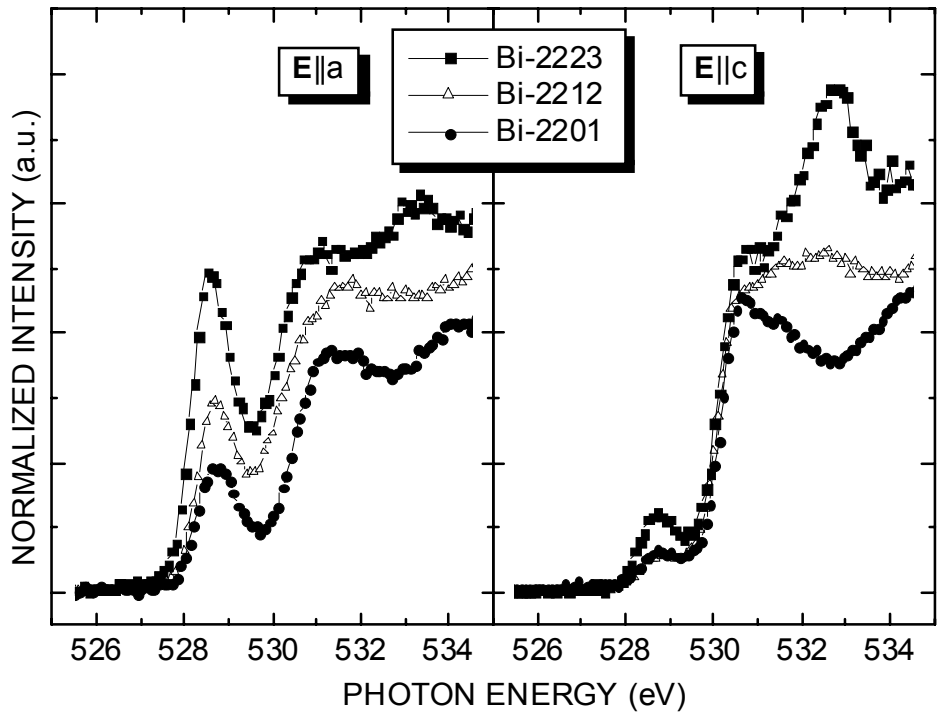

**Fig. 1** O 1s prepeak of $Bi_2Sr_{2-x}La_xCuO_{6+\delta}$ at optimal doping (x = 0.38) compared to Bi-2212 and Bi-2223 for two polarization geometries. The spectra are normalized at 600 eV photon energy. Left panel: polarization in the $CuO_2$ plane, right panel: polarization perpendicular to the $CuO_2$ plane.

In Fig. 1 the O 1s absorption spectrum of optimally doped n = 1-3 phase [8] is shown for two polarization geometries. For **E**∥a (normal incidence) the electrical field vector of the synchrotron radiation lies within the sample plane aligned parallel to the crystallographic a axis and for **E**∥c (grazing incidence) **E** is about 85% parallel to the surface normal. A single pre-peak at 528.7 eV can be seen in the normal incidence spectra of all three systems, corresponding to O 1s → O 2p transitions. Neglecting excitonic effects, one can conclude that oxygen-derived 2p final states are located at the Fermi level. They have the well-known x,y-symmetry due to the polarization dependence shown in Fig. 1 and have been viewed as empty states in a heavily p-doped O 2p-like valence band [1-4]. The main onset in the absorption at 530 eV photon energy corresponds to antibonding O 2p states. The intensities of the pre-peaks increase within experimental uncertainty linearly with n, the number of Cu-O layers per unit cell. Thus it is found that $T_C$ is not proportional to the height or intensity of the pre-peak since the $T_{C,max}$ of the phases is 29 K, 91 K, 108 K for n = 1, 2, and 3, respectively. Furthermore it is interesting to note that the n = 3-phase has a stronger contribution perpendicular to the $CuO_2$ planes.

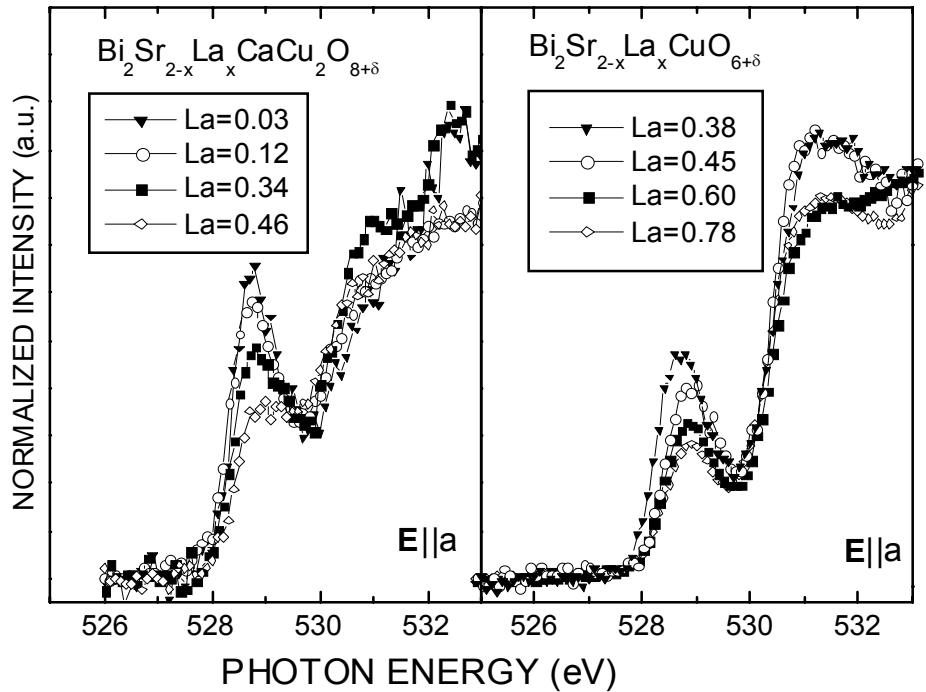

**Fig. 2** X-ray absorption spectra (XAS) of the O 1s prepeak of n = 2 $Bi_2Sr_{2-x}La_xCaCu_2O_{8+\delta}$ (left panel) for various doping levels (x = 0.03, 0.12, 0.34, 0.46) and XAS spectra of n = 1 $Bi_2Sr_{2-x}La_xCuO_{6+\delta}$ (right panel) for various doping levels (x = 0.38, 0.45, 0.60, 0.72). The $T_C$ of the n = 2 samples scales from 90 K (x = 0.03) down to 6 K (x = 0.43). The $T_C$ of the n = 1 samples scales from 29 K (x = 0.38) to 5 K (x = 0.72). The spectra are taken for **E**∥a and normalized at 600 eV photon energy.

In Fig. 2 we present first results of a La doped n = 2 system $Bi_2Sr_{2-x}La_xCaCu_2O_{8+\delta}$ (left panel) and La doped n=1 $Bi_2Sr_{2-x}La_xCuO_{6+\delta}$ system (right panel). For n = 2 the intensity of the pre-peak is highest for very low La doping (x = 0.03, $T_C$ = 90 K). With increasing La content both, the height of the pre-peak and the transition temperature decrease. In the case of n=1 lanthanum-free samples (x = 0) are found to be strongly hole overdoped, with a $T_C$ of about 7 K. In order to reach optimal $T_C$, part of the $Sr^{2+}$ must be replaced by $La^{3+}$ which reduces the hole concentration in the $CuO_2$ plane. Fig 2 (right panel) shows the evolution of the pre-peak when going from optimally (x = 0.38) to the hole-underdoped compound (x = 0.72). These experimental findings principally confirm the established view that the intensity of the pre-peak should dependent on the density of doped holes in the $CuO_2$ planes [1]. Nevertheless, the behaviour of overdoped Bi-2201 is much more intricating and will be studied in a forthcoming publication [9].

In summary, we found the intensity of the XAS pre-peak to scale with the number of $CuO_2$ layers per unit cell of the Bi-based cuprates, but not with $T_{C,max}$ of the respective samples. The hole density for the n = 3 phase showed a stronger out of plane-component. By La-substitution the density of doped holes could be varied continuously over a wide doping range for the n = 1 and n = 2 phase, giving an alternative to doping by Ca-substitution in between the $CuO_2$ layers.

This work was supported by the BMBF under project number 05 SB8 KH10.